\documentclass[showpacs,amsmath,amssymb,floatfix,prl,aps,twocolumn]{revtex4}
\usepackage{graphicx}
\usepackage{amssymb}

\def\sn{\textrm{sn}}
\def\cn{\textrm{cn}}
\def\dn{\textrm{dn}}
\def\cs{\textrm{cs}}
\def\sc{\textrm{sc}}
\def\K{\hbox{K}}
\def\real{\textrm{Re}}

\begin{document}

\title{Elliptic Phases: A Study of the Nonlinear Elasticity of Twist-Grain Boundaries}

\author{Christian D. Santangelo}
\author{Randall D. Kamien}
\affiliation{Department of Physics and Astronomy, University of Pennsylvania, Philadelphia, PA 19104-6396, USA}

\date{\today}

\begin{abstract}
We develop an explicit and tractable representation of a twist-grain-boundary phase of a smectic-$A$ liquid crystal.  This allows us to calculate the interaction energy between grain boundaries and the relative contributions from the bending and compression deformations.  We discuss the special stability of the $\pi/2$ grain boundaries and discuss the relation of this structure to the Schwarz D surface.
\end{abstract}


\maketitle
Topological defects are often the essential degrees of freedom.   They are the focus in the study of some phase transitions \cite{nelsonhalperin, YodhScience}, high-temperature superconductors \cite{bfn}, and liquid crystalline phases \cite{klemanbook}.   In the latter, there is a necessary connection between the topology of the defects and their geometry which is on the one hand theoretically challenging while, on the other hand, experimentally accessible through real-space, freeze-fracture imaging \cite{zas}.  The twist grain boundary (TGB) phase of smectic-A liquid crystals \cite{renn}, has an arrangement of screw dislocations which alter the geometry of the uniform, flat layers into a discretely rotating layered structure.  Though topology constrains the geometry, it does not  specify it.  Rather, the free energy of the deformed smectic layers  sets the periodicity of the lattice.  Prior analysis \cite{bluestein2} relied on linear elasticity to study small angle grain boundaries.  However, it has been shown that when the angles and deformations are large, the energetics of the rotationally-invariant nonlinear theory are not only quantitatively, but {\sl qualitatively} different than in the linear theory  \cite{kamien,bluestein,bps}.  In order to reconcile our understanding of the linear theory with the nonlinear elasticity and  in light of the recently observed \cite{clark} large angle grain boundaries, here we develop a full nonlinear theory of the largest angle grain boundaries allowed, with rotations of $\pi/2$.  To do this we explicitly sum the topological defects to render a closed-form expression {\sl which is an exact solution of the linear elasticity theory}.     Unlike parametric representations of surfaces, our surface is given as a multi-valued height function which allows us to directly calculate the compression energy and allows tractable comparison with real space images.  We directly calculate the energetics of space-filling TGB structures analytically and find that grain boundaries interact exponentially with separation. 

Smectic order is characterized by a periodic mass density $\rho = \rho_0 + \rho_1\cos\left[2\pi\Phi({\bf x})/a\right]$, where $\rho_0$ and $\rho_1$ are constant amplitudes, set at the nematic-to-smectic phase transition, and $a$ is the equilibrium layer spacing. The smectic layers are defined via the level sets of $\Phi$ through $\Phi({\bf x})/a=n\in\mathbb{Z}$, and the elastic free energy has two terms, 
\begin{figure}
\begin{center}
\resizebox{2.8in}{!}{\includegraphics{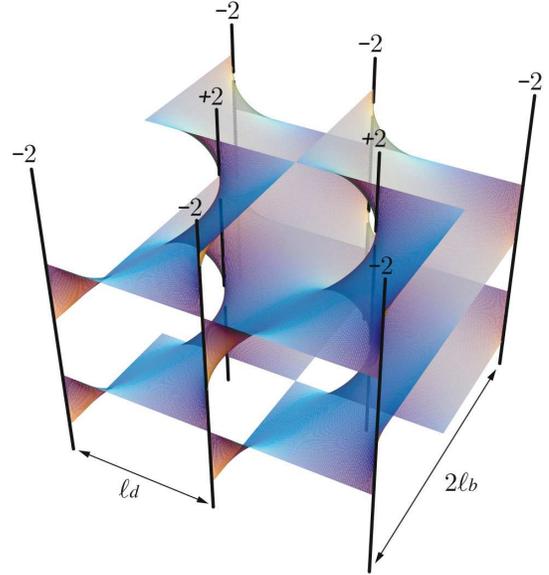}}
\caption{Schnerk's first surface, with charge $+2$ and $-2$ screw dislocations.  Note that the $-2$ dislocations lie at the center of the rectangle made by the adjacent $+2$ dislocations.   We choose $\theta=\psi$ and $k^2\approx  -0.03033$ so that $\K'(k)/\K(k) = 2-i$.  We are reminded that ``\ldots imaginary
things are often easier to see than real ones.'' \cite{Juster}}
\label{fig:schnerkpic}
\end{center}
\end{figure}
\begin{equation}\label{free}
F = \frac{B}{ 8}\int d^3\!x\left\{ \left[\left(\nabla\Phi\right)^2-1\right]^2 +  16\lambda^2 H^2\right\}
\end{equation}
where $B$ is the compression modulus, $\lambda^2\equiv K_1/B$, $K_1$ is the bending modulus, and $H\equiv\frac{1}{ 2}\nabla\cdot\left[\nabla\Phi/\vert\nabla\Phi\vert\right]$ is the mean curvature of the layers.
To study the energetics of defect-laden structures, we find an expression for 
the phase function $\Phi$ of the surface and use that to evaluate (\ref{free}).   For instance, a single screw dislocation at the origin is described by a helicoid~\cite{klemanbook} (here and throughout,  $x,y$, and $z$ are coordinates in $\mathbb{R}^3$ and $w\equiv x+iy$): 
\begin{equation}
\Phi_{\rm screw} = \gamma z -\frac{b}{ 2\pi}\arctan\left(\frac{y}{ x}\right)= \gamma z - \frac{b}{ 2\pi} \hbox{Im}\ln w 
\end{equation}
where the Burgers scalar $b/a$ must be an integer so that the mass density is single valued.  The prefactor $\gamma$ is necessary in order to allow $\vert\nabla\Phi\vert\rightarrow 1$  away from the defect.  For the screw dislocation this results in $\gamma^2=1-\lim_{w\rightarrow\infty} [b/(2\pi\vert w\vert)]^2=1$.  The helicoid is an extremal of (\ref{free}) as well as an
extremal of the often used quadratic free energy (which controls linear elasticity)
\begin{equation}
F=\frac{B}{2}\int d^3\!x\left\{ \left(\partial_z u\right)^2 + \lambda^2\left(\nabla_\perp^2 u\right)^2\right\}
\label{freeq}
\end{equation}
written in terms of $u\equiv z-\Phi$ and $\nabla_\perp\equiv \hat x\partial_x + \hat y\partial_y$.  Unlike the nonlinear theory, however, a screw dislocation has vanishing energy in the linear theory because $\Phi_{\rm screw}$ is a harmonic function of $w$, {\sl i.e.} $\nabla_\perp^2 \Phi_{\rm screw}=0$.

Do multiple screw dislocations interact?  In the linear theory (\ref{freeq}) there is no interaction.  The inclusion of 
non-Goldstone, nematic director modes results in long-distance exponential interactions as arise in flux lines \cite{renn}.  However, here we consider the interactions arising from the necessary {\sl nonlinearities} of a rotationally-invarient free energy (\ref{free}).   Our study proceeds by considering solutions of the linear theory and calculating their energy in the nonlinear theory.  Though these solutions are not minimizers of (\ref{free}) they have the desired topology of the smectic configurations of interest and, importantly, become exact in the limit of infinite separation between dislocations.  We begin with a single twist-grain boundary, a row of dislocations along the $x$-axis with uniform spacing $\ell_d$ . The phase field is a sum of individual helicoids:
\begin{equation}\label{eq:row}
\Phi_{\rm row} = \gamma z -\frac{b}{ 2\pi}\sum_{n=-\infty}^\infty \hbox{Im}\ln\left[w-n\ell_d \right]
\end{equation}
Utilizing the infinite product $\sin w = w\prod_{n\ne 0} \left(1-\frac{w}{ n\pi}\right)$ it follows that \cite{kamien2} (up to unimportant constants)
\begin{equation}
\Phi_{\rm row} = \gamma z-\frac{b}{ 2\pi}\hbox{Im}\ln\sin\left(\frac{\pi w}{ \ell_d}\right).
\end{equation}
Since the grain boundary lies on the $y=0$ axis, setting the compression strain to zero at $y=\pm \infty$  compels us to set $\gamma^2=1 -b^2/(2\ell_d)^2$.
The layer normal of a single grain boundary, ${\bf N}\equiv\nabla\Phi/\vert\nabla\Phi\vert$, rotates 
from ${\bf N}_-=[-\frac{b}{ 2\ell_d},0,\gamma]$ to ${\bf N}_+=[\frac{b}{ 2\ell_d},0,\gamma]$ as $y$ goes from $-\infty$ to $\infty$, implying a uniform rotation of the layers across the grain boundary by an angle $\sin \alpha = b \gamma/\ell_d$~\cite{renn,kamien}.

Because a rotation by $\pi$ amounts to no rotation at all, there is necessarily a dual description of a single grain boundary as a rotation of $\pi-\alpha$ by viewing the dislocations as parallel to the $x$-axis.  This dual description reflects the geometric nature of the defects and can be seen directly from the parametric equation for the level sets of $\Phi_{\rm row}$, $\tan(2\pi\gamma z/b)\tan(\pi  x/\ell_d)=\tanh(\pi y/\ell_d)$~\cite{kamien}.
Thus, after a rotation around the $y$-axis by $\pi/2$, $(x,z)\rightarrow (z,-x)$,  and the layer normal rotates from ${\bf N}_-=[\gamma,0,\frac{b}{2\ell_d}]$ to ${\bf N}_+=[-\gamma,0,\frac{b}{2\ell_d}]$.  The rotation angle becomes $\sin\alpha = -b\gamma/\ell_d$, for a total rotation of $\pi-\alpha$.
Note that the sense of the rotation has been reversed or, equivalently, $b\rightarrow-b$~\cite{kamien}.  

This duality is the lynchpin for our further analysis.  When piecing together single grain boundaries into a TGB phase, the defects in the grain boundaries must rotate with the smectic layers, pulling the topological defects along with the very geometry they create.  In the case of $\pi/2$ grain boundaries a special simplification occurs.  If the first grain boundary has defects along the $z$-axis, then the two adjacent boundaries should have defects along the $x$-axis. Employing the duality which swaps $x$ and $z$, we can choose to view the adjacent boundaries as made of defects along the $z$-axis with the {\sl opposite} Burgers scalar.  Thus, in the case of $\pi/2$ boundaries, we have a structure with only parallel screw dislocations along the the $z$ direction, with alternating signs!  The phase field for the sum of individual grain boundaries is
\begin{equation}\label{eq:schnerksum}
\Phi_{\rm TGB} =  \gamma z -\frac{b}{2 \pi}\hbox{Im} \sum_{m=-\infty}^\infty (-1)^m \ln\sin\left(\frac{\pi w}{ \ell_d} + m\frac{\pi\tau}{ 2}\right)
\end{equation}
where $\tau$ is a complex number which generates the appropriate translations along the $y$ direction.  Throughout we set $\tau \equiv  2 i \ell_b/\ell_d+1$, where $\ell_b$ is the distance between the grain boundaries \cite{renn}.  Note that this sum alternates, reflecting the alternating sign of the defects in adjacent boundaries.

The infinite sum in equation (\ref{eq:schnerksum}) can be put into closed form by observing that the exponential of the sum is doubly-periodic and, through rescaling of $x$ and $y$, shares all the poles and zeroes of the Jacobi elliptic function $\sn(u,k)$ \cite{elliptic}, or equivalently, through one of the established infinite products for $\sn(u,k)$.  This sum will generate a surface of the desired topology which is also a harmonic function and thus a minimizer of the quadratic free energy.  We arrive at the exact summation of screw dislocations for
a $\pi/2$ TGB structure:
\begin{equation}\label{eq:infiniteproduct}
\Phi_{\rm TGB}=\gamma z - \frac{b}{2 \pi}\hbox{Im}\ln \sn \left[\theta x + i \psi y,k\right]
\end{equation}
where $\theta \equiv 2 \K(k)/\ell_d$,  $\psi \equiv \real \K'(k)/\ell_b$ are scale factors, $\K(k)$ is the complete elliptic integral of the first kind, $\K'(k) \equiv \K(\sqrt{1-k^2})$, and $k$ is the elliptic modulus, which for our purposes must be pure imaginary ({\sl i.e.} $k^2$ is real and nonpositive)~\cite{ellipticfootnote},.   The level sets of $\Phi_{\rm TGB}$, dubbed Schnerk's first surface, are shown in Fig.~\ref{fig:schnerkpic} for $ \ell_b = \ell_d$.  To simplify our notation, we define $\zeta = \theta x  + i\psi y$, use Glaisher's notation ($\cs$ for $\cn/\sn$, {\sl etc.}) for the elliptic functions \cite{elliptic}, and suppress the elliptic modulus $k$.  

The orthorhombic symmetry of Schnerk's surface suggests that, like the single grain boundary, it can be viewed as one of three orthogonal arrangements of screw defects.  To see this, we note that the level sets satisfy
\begin{equation}\label{eq:parametricform}
\tan(2 \pi \gamma z/b) \frac{\sc[\theta x]}{\dn[\theta x]} = -i\frac{\sc[i\psi y]}{\dn[i\psi y]}
\end{equation}
We recognize $\sc(\zeta)/\dn(\zeta)$ as the elliptic generalization of $\tan\zeta$.  By considering the zeroes and poles of the elliptic functions, it is possible to view the surface as  being composed of oppositely charged, staggered defects along $x$ or $y$, instead of $z$, generalizing the duality of a single grain boundary \cite{kamien}. 

The closed-form expression for $\Phi_{\rm TGB}$ allows us to study the energetics of this structure by use of the established properties of elliptic functions.  For instance, the compression strain $u_{zz}\equiv [1-(\nabla \Phi)^2]/2$ is the somewhat cumbersome
\begin{eqnarray}
u_{zz}&=& \frac{1}{2} \Big[1 - \gamma^2  - \frac{b^2}{8 \pi^2} \left(\theta^2 + \psi^2\right) \left\vert\cs\,\zeta\; \dn\,\zeta\right\vert^2\nonumber\\
& & +\frac{b^2}{8 \pi^2} \left(\theta^2 - \psi^2\right) \hbox{Re}\left[ \cs^2\zeta\; \dn^2\zeta \right]\Big]
\end{eqnarray}
This expression demonstrates that we can choose the pair $(\theta,\psi)$ in order to simplify our analysis; because we can set the periodicity of the lattice by altering $(\theta,\psi)$ or $k$, we can freely set the pair at our convenience.  Though these are distinct deformations of the layered structure they share the same periodicity and topology.  We will focus on the case $\psi=\theta$ in the following, though our results do not change qualitatively for other choices.  The elliptic modulus is set by $\textrm{Re} \K'(k)/\K(k) = 2 \ell_b/\ell_d$.
Recall that in the case of a single grain boundary, we set $\gamma$ by considering $y=\pm\infty$.  Here, the structure is triply periodic and there is no ``infinity''.  We choose instead to have the compression vanish halfway between the grain boundaries, {\sl e.g.} along $y=\ell_b/2$ or $x=\ell_d/4$.  Because these lines are where we would measure the rotation of the layers, this is a natural choice and, for $k^2<0$, the compression strain is constant along these lines.  This allows us to set both $\gamma$ and $\ell_d$.  Though in principle we should choose $\gamma$ to minimize the compression energy for a single periodic domain, as $\ell_b/\ell_d\rightarrow\infty$ the latter procedure will yield our solution.
 
We make use of the following  expansion \cite{jacobi} in terms of $q\equiv \exp\left[-\pi\K'/\K\right] = -\exp\left[-2\pi \ell_b/\ell_d\right]$:
\begin{equation}\label{asym}
\ln\left[\frac{\sqrt{k}\;\sn\,\zeta}{2 q^{1/4}\sin\left(\pi w/ \ell_d\right)}\right] = \sum_{m=1}^\infty \frac{2}{ m}\frac{q^m}{ 1+q^m} \cos\left(\frac{2m\pi  w}{\ell_d}\right)
\end{equation}
The utility of this expression is that it and its derivative have, as their leading terms, the trigonometric functions present in the single grain boundary (\ref{eq:row}), allowing us to isolate the energetic corrections arising from interactions.  To calculate the interaction energy between grain boundaries, we expand the free energy in powers of $q$. Since $q=-\exp\left[-2\pi\ell_b/\ell_d\right]$, the interactions will fall off {\sl exponentially} with $\ell_b$, our central result.  We note that this conclusion is independent of our choices of $\theta$ and $\psi$ as long as we are in a regime where $q$ is small.
For large $\ell_b/\ell_d$, $q$ is small making this a good expansion.  Note that even for the symmetric case where $\ell_b=\ell_d$, shown in Fig. 1, we have $q=-e^{-2\pi}\approx -0.002$, small enough to use (\ref{asym}) reliably.     We compare this to the interaction between defects in a linear theory with a director field: in the latter case the decay length is $\lambda$, not
$\ell_d$.  The interaction in the nonlinear theory arises from long-distance strain not from 
the decay of gauge-like director modes.

In order to find the interaction between grain boundaries, we expand the energy per unit area in terms of $q$ and remove the energy of $L_y/\ell_b$ isolated grain boundaries, where $L_y$ is the $y$ dimension of the system.  The compression interaction per area $A$ is proportional to $q/\ln q$ and we find:
\begin{equation}\label{interact}
\frac{\Delta F_c}{A}\sim \frac{BL_z\ell_d}{2\pi\ell_b}\left[C+\left(\frac{b}{2\pi\xi}\right)^2 +2\ln\left(\frac{2\sqrt{2}\xi}{b}\right)\right]q
\end{equation}
where $C$ is a positive constant of order unity, $L_z$ is the $z$-dimension of the system, and an elastic cutoff length $\xi$ is introduced to cutoff a $\vert w\vert^{-4}$ divergence in $u^2_{zz}$ near the origin.   This cutoff is necessary in the case of a single dislocation and a single grain boundary \cite{kamien} as well.   Since $q<0$, we find an attractive, exponential interaction.  The attraction comes as no surprise since the adjacent grain boundaries are made of {\sl opposite} signed dislocations.   Note that the constant term in (\ref{interact}) gets a contribution from the ``tails'' of the single grain boundaries and is independent of both the Burgers scalar and the cutoff.  The numerical details of (\ref{interact}) depend on the geometry of the cores, but our result is representative of the energetics.  Minimizing over $\xi$ gives $\xi\propto b$ as in nonlinear theories of edge dislocations \cite{klemanbook,bps}.

\begin{figure}[t]
\begin{center}
\resizebox{2.8in}{!}
{\includegraphics{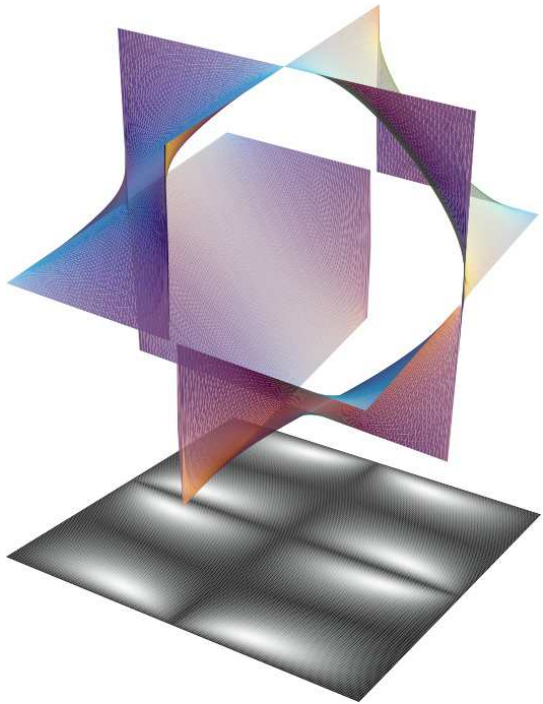}}
\caption{The unit cell of Schnerk's surface shares the topology of the Schwarz D surface.  However, Schnerk's surface is not minimal.  We project the value of $\vert H\vert$, the magnitude of the mean curvature of the surface, in grayscale onto the floor (black is zero curvature, white is the maximum curvature).}
\label{fig:Dcell}
\end{center}
\end{figure}

The bending energy does not require a cutoff, as the mean curvature is finite everywhere (see Fig. \ref{fig:Dcell}).  This energy may also be expanded in powers of $q$  and we would, again, find an exponential interaction.   In the two special cases of $\psi=\sqrt{2}\theta$ for any $k$ and $\theta=\psi$ with $k^2=-1$, the interaction between grain boundaries is purely repulsive, the first case recapitulating the vanishing of $H$ for Scherk's first surface \cite{kamien}.  Whether the curvature interaction is always repulsive is under investigation \cite{tobe}.

Our construction demonstrates that the $\pi/2$ TGB structure is, in fact, charge neutral from the point of view of the screw dislocations.  Indeed, any TGB structure with rotation angle $\alpha=\pi/n$ for $n\in\mathbb{Z}$ will be charge neutral -- the screw dislocations in one boundary will have equal and oppositely charged defects in the $n^{\rm th}$ further grain boundary.  However, the $\pi/2$ case is special because the cancellation occurs in the {\sl adjacent} grain boundary which suggests that the $\pi/2$ structure is especially stable.   Though the structure we have studied here is not chiral, a nematic director can rotate uniformly through the structure, pulling away from the surface normal as necessary.  For ${\bf n}=[\cos(\frac{\pi y}{2\ell_b}),0,\sin(\frac{\pi y}{2\ell_b})]$, we find the average alignment of the director with the layer normal to be $S=\frac{3}{2}\langle({\bf N}\cdot{\bf n})^2\rangle-\frac{1}{2}\approx 0.7$ for the geometry shown in Fig. 1.  This value suggests that the director and the layer normal are more or less aligned.   If the geometry of the mesogens were to favor positive saddle-splay or, equivalently, negative Gaussian curvature then this would further stabilize this phase over flat layers.  It may be possible to study grain boundaries with rotation angles $\pi/n$ by grouping 
blocks together into ``superblocks'' which interact as effective $\pi/2$ grain boundaries.  Whether
this can be made precise, even perturbatively in $q$, is under investigation \cite{tobe}.

In closing, we note that the unit cell of Schnerk's first surface shown in Fig.~\ref{fig:Dcell} has the same bicontinuous topology as the minimal Schwarz D surface.   In Fig.~\ref{fig:Dcell} we have also indicated the magnitude of the mean curvature, $\vert H\vert$, which does not vanish everywhere.   We recall that a single grain boundary can be made minimal by a stretch of $\sec(\alpha/2)$ along the $y$-axis to become Scherk's first surface \cite{kamien}.   A possible generalization, in the spirit of recent solutions of maximal surfaces in Minkowski space \cite{Hoppe},  is to replace $\tan(2 \pi\gamma z/b)$ with its elliptic analog $\sc(2\pi\gamma z/b)/\dn(2\pi\gamma z/b)$ in (\ref{eq:parametricform}) to achieve greater symmetry between the $x,y$, and $z$ coordinates.  Similarly, the parametric Weierstra\ss-Enneper representation \cite{nitsche} of the coordinates of a minimal surface can, in some cases, be reduced to elliptic functions \cite{weellip}.  We are unaware of choices of $\theta$, $\psi$, or $k$ or other simple deformations that make the curvature of Schnerk's surface constant.  

We have directly summed the phase fields for an infinite array of screw dislocations which generates the
topology of $\pi/2$ TGB phase and which is a solution of the linear elasticity theory.  This provided an analytically tractable representation of the structure which we used to study the interaction energy between grain boundaries in a rotationally-invariant elasticity.  We find exponential interactions -- the true minimizer should have interactions at least as weak.  Further work will explore additional deformations, the character of the cutoff $\xi$, and make intimate comparison with experiment \cite{clark}.  

\begin{acknowledgments}
We thank M. Cohen, G. Gibbons, R. Kusner, and T.C. Lubensky for discussions. This work was supported through NSF Grants DMR01-29804 and DMR05-47230, the Donors of the ACS Petroleum Research Fund, and a gift from L.J. Bernstein.
\end{acknowledgments}


\begin{thebibliography}{99}


\bibitem{nelsonhalperin}
B.I.~Halperin and D.R.~Nelson, Phys. Rev. Lett. {\bf 41}, 121 (1978);
D.R.~Nelson
and B.I.~Halperin, Phys. Rev. B {\bf 19}, 2457 (1979) 2457.

\bibitem{YodhScience} A.M. Alsayed, {\sl et al.}, Science {\bf 19}, 1207 (2005).

\bibitem{bfn} L. Balents, M.P.A. Fisher, and C. Nayak, Int. J. Mod. Phys. B {\bf 12}, 1033 (1998).

\bibitem{klemanbook} M. Kleman, \textit{Points, Lines, and Walls} (Wiley, New York, 1983).


\bibitem{zas} K.J. Ihn, {\sl et al.}, Science {\bf 258}, 275 (1992).



\bibitem{renn} S.R. Renn and T.C. Lubensky, Phys. Rev. A \textbf{38}, 2132 (1988).

\bibitem{bluestein2} I. Bluestein, R.D. Kamien, and T.C. Lubensky, Phys. Rev. E \textbf{63}, 061702 (2001).  

\bibitem{kamien} R.D. Kamien and T.C. Lubensky, Phys. Rev. Lett. \textbf{82}, 2892 (1999).

\bibitem{bps} C.D. Santangelo and R.D. Kamien, Phys. Rev. Lett. \textbf{91}, 045506 (2003); Proc. R. Soc. A {\bf 461}, 2911 (2005).

\bibitem{bluestein} I. Bluestein and R.D. Kamien, Europhys. Lett. \textbf{59}, 68 (2002).

\bibitem{clark} J. Fernsler, {\sl et al.}, Proc. Natl. Acad. Sci. USA \textbf{102}, 14191 (2005).

\bibitem{Juster} N. Juster, {\sl The Phantom Tollbooth}, (Epstein \& Carroll, New York, 1961).

\bibitem{kamien2} R.D. Kamien, Appl. Math. Lett. \textbf{14}, 797 (2001).

\bibitem{elliptic} H. Hancock {\sl Lectures on the Theory of Elliptic Functions} (Dover, New York, 2004).

\bibitem{ellipticfootnote} The elliptic integral integral of the first kind is $\K(k) = \int_0^1dx [(1-x^2) (1- k^2 x^2)]^{-1/2}$.  When $k$ is pure imaginary, $\textrm{Im}{\K'(k)} = - \K(k)$. 

\bibitem{jacobi} C.G.J. Jacobi, {\sl Fundamenta nova theoriae functionum ellipticarum}, (Regiomonti, Sumtibus fratrum Borntraeger, 1829),  \S\S 39, Eq. (6).

\bibitem{tobe} C.D. Santangelo and R.D. Kamien, {\sl unpublished}.

\bibitem{Hoppe} J. Hoppe, hep-th/9503069; G.W. Gibbons, Nucl. Phys. B {\bf 514}, 603 (1998)

\bibitem{nitsche} J.C.C.~Nitsche, {\sl Lectures on Minimal
Surfaces},
(Translated by J.M.~Feinberg), (Cambridge University Press, Cambridge, 1989).


\bibitem{weellip} D. Cvijovi\'c and J. Klinowski, J. Phys. I France {\bf 2}, 137 (1992); {\bf 2}, 2191 (1992); {\bf 2}, 2207 (1992).





\end{thebibliography}
\end{document}